\documentclass[conference]{IEEEtran}
\IEEEoverridecommandlockouts

\usepackage{cite}
\usepackage{amsmath,amssymb,amsfonts}
\usepackage{algorithmic}
\usepackage{hyperref}
\usepackage{graphicx}
\usepackage{multirow}
\usepackage{bbding}
\usepackage{textcomp}
\usepackage{makecell}
\usepackage{array}
\usepackage{booktabs}
\usepackage{pgfmath}
\usepackage{utfsym}
\usepackage{fontawesome}

\usepackage{cite}
\usepackage{algorithm}
\usepackage[numbers,sort&compress]{natbib}
\usepackage{comment}
\usepackage{subfigure}
\usepackage{arabtex}
\usepackage{adjustbox}
\usepackage{caption}
\usepackage{subcaption}

\usepackage{multicol}
\usepackage{color, colortbl}
\colorlet{eng}{blue!10}
\colorlet{cmn}{teal!10}
\colorlet{multi}{yellow!10}
\colorlet{euro}{orange!10}

\usepackage{pifont} 
\definecolor{bsRed}{rgb}{0.95, 0.0, 0.0}

\def\BibTeX{{\rm B\kern-.05em{\sc i\kern-.025em b}\kern-.08em
    T\kern-.1667em\lower.7ex\hbox{E}\kern-.125emX}}
\begin{document}

\title{Ultra-Low Latency Speech Enhancement - \\A Comprehensive Study 
\thanks{Work performed during H. Wu's internship at Microsoft Research.}
}

\author{
Haibin Wu$^1$, Sebastian Braun$^2$ 
\\
\textit{$^1$National Taiwan University},    \textit{$^2$Microsoft, Redmond, WA, USA}
}

\maketitle

\begin{abstract}
Speech enhancement models should meet very low latency requirements typically smaller than 5 ms for hearing assistive devices. 
While various low-latency techniques have been proposed, comparing these methods in a controlled setup using DNNs remains blank. 
Previous papers have variations in task, training data, scripts, and evaluation settings, which make fair comparison impossible. 
Moreover, all methods are tested on small, simulated datasets, making it difficult to fairly assess their performance in real-world conditions, which could impact the reliability of scientific findings.
To address these issues, we comprehensively investigate various low-latency techniques using consistent training on large-scale data and evaluate with more relevant metrics on real-world data. 
Specifically, we explore the effectiveness of asymmetric windows, learnable windows, adaptive time domain filterbanks, and the future-frame prediction technique. 
Additionally, we examine whether increasing the model size can compensate for the reduced window size, as well as the novel Mamba architecture in low-latency environments.
\end{abstract}


\section{Introduction}
\label{sec:intro}

Hearables and wearable audio devices have gained popularity recently, making low-latency and real-time scenarios increasingly crucial in speech enhancement (SE).
Popular short-time Fourier transform (STFT)-based SE models \cite{braun2021towards,tan2019learning} usually employ a 20 ms window for STFT / inverse STFT with a 10 ms hop. 
The total latency is the window length (20 ms), a summation of algorithmic latency (10 ms) due to the overlap-add method used in signal re-synthesis, and the buffer latency (hop length 10 ms). 
Processing time for each frame is not included.
However, the audible threshold for hearables, where the direct sound may leak through and mix with the processed sound, is less than 5 ms \cite{graetzer2021clarity}. 
In addition to hearing assistive devices, reducing the total latency of voice over Internet protocol (VoIP) pipelines \cite{goode2002voice} by minimizing the latency of each processing block is another key application. 

While several low-latency SE techniques have been proposed over the last decade, a controlled and systematic comparison of these methods, especially using modern DNN models, is seldom investigated. 
\cite{wang2021deep} use asymmetric windows \cite{mauler2007low} with a by now outdated architecture for speech separation, with only a single 8~ms window setting, evaluate only on small scale synthetic data and do not compare to other methods. \cite{luo2019conv,maciejewski2020whamr,casebeer2020efficient} use learnable transforms, but do not focus on the low latency aspect, therefore not comparing to other methods and mostly evaluate on small scale synthetic data.
\cite{zheng2022low} proposes an interesting filterbank equalizer (FBE), but do not compare to obvious baselines like STFT domain algorithms with asymmetric windows. 
A future frame prediction technique is proposed in \cite{iotov2022computationally,wang2022stft} to reduce latency, but the key deciding baseline comparing a model with and without prediction at \textit{same} latency is not provided.
The Clarity Challenge
\cite{cox2023overview} evaluates only intelligibility, not audio quality, and rather compares overall systems from entirely different pipelines, whereas we investigate details by swapping out single modules in one base system, which provides detailed insights.

There are two key challenges to fairly compare all low-latency techniques in real-world SE:
(a). Different low-latency tricks are implemented in different settings. Even minor differences in training hyper-parameters can lead to significantly different results \cite{watanabe2018espnet,yang2024large}, not to mention inconsistencies in evaluation settings and training data.
(b). Most works present results only on small, simulated datasets, which sometimes can transfer surprisingly different to real-world conditions. Some trends or advantages observed in small-scale datasets may not hold when applied to large-scale, real-world data.
Our main contributions are summarized as follows:
(1). We implement all models in a unified framework to exclude impacts from different training data and pipelines, model architectures, loss functions, and evaluation settings. 
(2). We evaluate on large-scale data and more precise state-of-the-art metrics to allow direct conclusions relevant to real-world performance in commercial applications.
(3). This is the first work to fairly evaluate five different low latency techniques, including STFT based models with symmetric and asymmetric windows, directly learning transforms from time domain, FBE (adaptive time domain filtering) and the future frame prediction.

\section{Base enhancement pipeline}
\label{subsec:base}

Single-channel SE aims to estimate the clean signal 
given the noisy waveform $x(t) = s(t) + n(t)$, where
$n(t)$ combines the added and reverberant noise.

Our base enhancement pipeline comprises three processing blocks:
1). The analysis transform transforms short segments of the noisy waveform into time-frequency domain representations.
2). The SE model then processes these representations to produce enhanced versions at each time step. 
3). The synthesis transform converts the enhanced representations into enhanced waveform $\hat{s}(t)$.
The objective is to make $\hat{s}(t)$ a faithful estimation to $s(t)$.

\noindent
\textbf{Analysis transform}: 
We divide the input mixture waveform $x(t)$ into overlapped segments with length $L_\text{a}$ and $P$ samples frame shift. 
Each segment is represented by $\mathbf{x_k} \in \mathbb{R}^{1\times L_\text{a}}$, where $k = 1,\dots,K$ is the segment index and $K$ is the total number of segments. 
Each segment $\mathbf{x}_k$ is transformed into a $N$-dimensional representation, $X_k \in \mathbb{R}^{1\times N}$ via a 1-D convolution, detailed as:
\begin{align}
X_k = (\mathbf{x_k} \circ \mathbf{w}_\text{a}) \cdot W_{fft}
\label{eqn:enc}
\end{align}
where $W_{fft} \in \mathbb{R}^{L_\text{a}\times N}$ contains $N$ vectors (analysis transform basis functions, and can be initialized as Fourier transform basis functions) with length $L_\text{a}$ each, $\cdot$ denotes matrix multiplication, $\circ$ denotes element-wise product, $\mathbf{w}_\text{a} \in \mathbb{R}^{1 \times L_\text{a}}$ is the analysis window.
After the analysis transform processing, we can get the encoded time-frequency domain representations $X \in \mathbb{R}^{N \times K}$.
The combination of window $\mathbf{w}_\text{a}$ and transform matrix $W_{fft}$ can be implemented as a 1-D convolutional layer.

\noindent
\textbf{The enhancement model} 
enhances $X \in \mathbb{R}^{N \times K}$ through $\hat{S} = f_{\theta} (X)$:
where $f_{\theta}$ denotes the DNN model (predicting either enhancement filters or direct signal mapping) parameterized by $\theta$, and $\hat{S} \in \mathbb{R}^{N \times K}$ is the enhanced version of $X$.

\noindent
\textbf{Synthesis transform:} 
The synthesis transform reconstructs the waveform from each segment of the enhanced representation $\hat{X}$.
For example, given $\hat{S}_{k} \in \mathbb{R}^{1 \times \hat{N}}$, where $k$ denotes the segment index, the synthesis transform is performed by:
\begin{align}
\mathbf{\hat{s}_{k}} = (\hat{S}_{k} \cdot W_{ifft}) \circ \mathbf{w}_\text{s}
\label{eqn:dec}
\end{align}
where $\mathbf{\hat{s}}_{k} \in \mathbb{R}^{1\times L_s}$ is the transformed waveform, $W_{ifft} \in \mathbb{R}^{N\times L_s}$ contains $N$ vectors (synthesis transform basis functions) with length $L_s$ each, $\mathbf{w}_\text{s} \in \mathbb{R}^{1\times L_s}$ denotes the synthesis window. 
The overlapping segments are summed together to generate the final enhanced waveform $\hat{s}(t)$. 
The overlap-add process including windowing given by \eqref{eqn:dec} can be implemented as a single transpose convolutional layer.
Note that the overlap-add procedure introduces an algorithmic latency of $L_\text{s}-P$.

\section{Low latency processing methods}
\label{subsec:latency}
In this section, we introduce low-latency strategies from previous studies and integrate them into our unified base model to allow as fair a comparison as possible.
As backbone SE model ($f_{\theta}$), we use the Convolutional Recurrent U-Net for Speech Enhancement (CRUSE) \cite{braun2021towards} due to its balanced performance/compute footprint. 
It consists of a series of convolutional layers, a 4-channel GRU bottleneck, mirrored deconvolutional layers and conv-skip connections. 
We enhance via causal 3x3 deep filters as described in \cite{braun2022,mack2019deep}. 
Typically, time-frequency processing techniques use symmetric, i.\,e.\, identical analysis and synthesis window pairs with $L_\text{a} = L_\text{s}$ as they are easy to design for perfect reconstruction \cite{hansler2008speech} and well-behaved. Here we use sqrt-Hann windows.
Note that for fair comparison, we keep the model architecture consistent, independent of the window size. 
Also for shorter windows, the architecture is identical to the 20 ms window one, but when reducing window sizes, we keep the frequency resolution $N$, i.\,e.\ feature size, constant by zero-padding the windows.

\subsection{Asymmetric windows in analysis and synthesis transforms} 
As the latency of overlap-add algorithms is only determined by the synthesis window length, shortening the synthesis window reduces latency, while the analysis window length can be kept to not harm frequency resolution \cite{mauler2007low,wang2021deep}.
We use the asymmetric window pair proposed in \cite{mauler2007low,wang2021deep}, which consists of the concatenation of two half Hann windows of different lengths for analysis, and a shorter Hann window for synthesis.


\subsection{Learnable analysis and synthesis transforms} 
\cite{luo2019conv,maciejewski2020whamr,casebeer2020efficient} investigate learnable analysis and synthesis transforms, partially demonstrating performance improvements over fixed analysis and synthesis transforms. 
Specifically, they use trainable convolution / transposed convolution layers, or separate trainable matrices to fulfill \eqref{eqn:enc} and \eqref{eqn:dec}.
\textit{However, all studies \cite{luo2019conv,maciejewski2020whamr,casebeer2020efficient} are restricted to symmetric window settings and do not explore asymmetric settings. 
\cite{luo2019conv,maciejewski2020whamr} focus on speech separation. 
Further, the advantages shown on non-reverberant or synthetic datasets sometimes vanish when applied to real-world and reverberant cases.}
We implement the trainable version of asymmetric analysis and synthesis transforms by using trainable convolution / transposed convolution layers for Equation~\eqref{eqn:enc} and \eqref{eqn:dec}. 
We further found adding a ReLU nonlinearity at the analysis improves performance and stability.

\subsection{Trainable filterbank equalizer}
To heavily reduce latency, time-domain processing techniques using short finite impulse response (FIR) filters seem an attractive choice. 
The filterbank equalizer (FBE) proposed in \cite{lollmann2007uniform} has been adopted for a DNN in \cite{zheng2022low}. 
This adaptive FBE \cite{zheng2022low} predicts a bank of $M$ time-variant short FIR filters of length $2P$ for each frame. 
It was designed as a two-stage architecture to first predict longer filters and a second LSTM-based stage to shorten the filters for low latency. 
To maintain fair comparability, we integrate the filter shortening module in our base CRUSE model by letting the last convolutional layer predict directly the set of $2M$ (double because complex) filters of length $N$, which are then shortened by a linear mapping layer from $N$ to $2P$. 
After filtering is carried out as complex multiplication with a FFT of the corresponding input chunk, the output time frame is obtained by iFFT and keeping the last $P$ samples (overlap-discard).
Now the total latency of this system is determined only by the length of these filters, i.\,e.\, the hop-size $P$.
The resulting model size and complexity change only marginally and remain, therefore, comparable.

\subsection{Future frame prediction}
To reduce the latency, \cite{iotov2022computationally,wang2022stft} proposed to predict one future frame ahead, which seems reasonable when using overlapping frames, as a portion of the future information is already available in the analysis window.
Specifically, this technique predicts the future enhanced frame $\hat{S}_{k+1}$ from observations $X_0, X_1, ..., X_{k}$ only up to frame $k$. 
Our model uses a mapping-based target \cite{xu2014regression} to predict the enhanced features, specifically, our model predicts the complex compressed STFT spectrum.
Predicting one frame ahead reduces the latency by one hop-size, leading to zero algorithmic latency for 50\% overlap. 
The resulting total latency is then only the buffer latency (hop size). 
While (deep) filtering \cite{mack2019deep} typically outperforms mapping \cite{xu2014regression} for SE, the future-frame prediction model can only use a mapping-based target, as a filtering-based target requires the current frame for enhancement, which may be a disadvantage.

Notable \textbf{missing settings} in prior works investigating future frame prediction are: 
(1) they only use mapping based systems, but we don't know if prediction with mapping can still be useful when compared to more performant filtering based methods at same latency. 
(2) \cite{wang2022stft} demonstrates that predicting one frame ahead only slightly degrades performance by comparing scenarios with and without future-frame prediction under the same window size, which however results in different practically relevant \emph{total latency} of these two systems.
In this paper, we investigate these two missing settings in the context of large-scale, real-world speech enhancement.

\renewcommand{\arraystretch}{0.8}
\begin{table*}[ht]
\centering
\caption{Comparison for different windows and models. iWin and oWin denote the input and output window lengths.}
\vspace{-8pt}
\label{tab:all}
\adjustbox{max width=0.95\textwidth}{
\begin{tabular}{cc|ccc|cc|ccc|ccccccc}
\toprule
\multirow{2}{*}{\textbf{Row}} & \multirow{2}{*}{\textbf{Model Name}} & \multirow{2}{*}{\textbf{iWin}} & \multirow{2}{*}{\textbf{oWin}} & \multirow{2}{*}{\textbf{Latency}} & \multirow{2}{*}{\textbf{Model Size}} & \multirow{2}{*}{\textbf{MACs}} & \multicolumn{3}{c}{\textbf{DNS blind set}} & \multicolumn{6}{c}{\textbf{simulated data}} \\
\cmidrule(lr){8-10} \cmidrule(lr){11-16}
 &  & & & & & & \textbf{SIG} & \textbf{BAK} & \textbf{OVR} & \textbf{SIG} & \textbf{BAK} & \textbf{OVR} & \textbf{PESQ} & \textbf{STOI (\%)} & \textbf{SISDR} \\
\midrule
A1 & sym-20ms & 20 & 20 & 20 & 0.625M & 230.27M & 3.18 & 3.78 & 2.79 & 2.87 & 4.03 & 2.46 & 2.22 & 84.45 & 10.52 \\
A2 & sym-10ms & 10 & 10 & 10 & 0.625M & 460.53M & 3.16 & 3.81 & 2.79 & 2.83 & 4.03 & 2.43 & 2.24 & 84.59 & 10.32 \\
A3 & sym-5ms & 5 & 5 & 5 & 0.625M & 921.06M & 3.12 & 3.77 & 2.74 & 2.78 & 4.02 & 2.38 & 2.21 & 84.46 & 9.94 \\
A4 & sym-3ms & 3 & 3 & 3 & 0.625M & 1.54G & 3.12 & 3.77 & 2.74 & 2.76 & 4.02 & 2.36 & 2.18 & 84.23 & 9.65 \\
\midrule
B1 & asym-10ms & 20 & 10 & 10 & 0.625M & 460.53M & 3.15 & 3.78 & 2.77 & 2.84 & 4.03 & 2.44 & 2.22 & 84.34 & 10.18 \\
B2 & asym-5ms & 20 & 5 & 5 & 0.625M & 921.06M & 3.13 & 3.77 & 2.75 & 2.81 & 4.03 & 2.41 & 2.16 & 84.20 & 10.06 \\
B3 & asym-3ms & 20 & 3 & 3 & 0.625M & 1.54G & 3.15 & 3.78 & 2.77 & 2.80 & 4.02 & 2.40 & 2.16 & 84.17 & 9.85 \\
\midrule
C1 & Learn-asym-10ms & 20 & 10 & 10 & 0.625M & 460.53M & 3.23 & 3.78 & 2.83 & 2.88 & 4.03 & 2.47 & 2.31 & 84.88 & 10.84 \\
C2 & Learn-asym-5ms & 20 & 5 & 5 & 0.625M & 921.06M & 3.19 & 3.81 & 2.82 & 2.80 & 4.03 & 2.40 & 2.31 & 85.02 & 10.64 \\
C3 & Learn-asym-3ms & 20 & 3 & 3 & 0.642M & 1.54G & 3.14 & 3.78 & 2.76 & 2.76 & 4.02 & 2.37 & 2.22 & 84.22 & 10.13 \\
\midrule
D1 & asym-same\_compute-10ms & 20 & 10 & 10 & 0.551M & 249.8M & 3.13 & 3.80 & 2.76 & 2.82 & 4.02 & 2.41 & 2.18 & 84.08 & 9.90 \\
D2 & asym-same\_compute-5ms & 20 & 5 & 5 & 0.157M & 235.04M & 3.08 & 3.71 & 2.68 & 2.76 & 4.01 & 2.36 & 2.10 & 83.04 & 9.29 \\
D3 & asym-same\_compute-3ms & 20 & 3 & 3 & 0.142M & 230.33M & 3.09 & 3.66 & 2.67 & 2.74 & 3.99 & 2.33 & 2.05 & 82.77 & 9.03 \\
\midrule
E1 & Mamba-20ms & 20 & 20 & 20 & 0.642M & - & 3.19 & 3.80 & 2.81 & 2.87 & 4.03 & 2.46 & 2.27 & 84.59 & 10.61 \\
E2 & Mamba-10ms & 10 & 10 & 10 & 0.642M & - & 3.10 & 3.77 & 2.72 & 2.81 & 4.02 & 2.40 & 2.21 & 83.96 & 10.16 \\
E3 & Mamba-5ms & 5 & 5 & 5 & 0.642M & - & 3.03 & 3.67 & 2.63 & 2.73 & 4.00 & 2.33 & 2.15 & 83.42 & 9.71 \\
\midrule
F1 & asym-10ms-medium & 20 & 10 & 10 & 2.32M & 1.09G & 3.19 & 3.77 & 2.80 & 2.85 & 4.02 & 2.44 & 2.30 & 84.75 & 10.56 \\
F2 & asym-5ms-medium & 20 & 5 & 5 & 2.32M & 2.20G & 3.16 & 3.80 & 2.78 & 2.84 & 4.03 & 2.43 & 2.19 & 84.15 & 10.07 \\
F3 & asym-3ms-medium & 20 & 3 & 3 & 2.32M & 3.66G & 3.18 & 3.78 & 2.80 & 2.82 & 4.02 & 2.42 & 2.22 & 84.68 & 10.16 \\
F4 & asym-10ms-large & 20 & 10 & 10  & 8.67M  & 3.28G & 3.22 & 3.80 & 2.84 & 2.87 & 4.04 & 2.47 & 2.33 & 85.24 & 10.70 \\
F5 & asym-5ms-large & 20 & 5 & 5 & 8.67M & 6.57G & 3.22 & 3.77 & 2.83 & 2.85 & 4.03 & 2.45 & 2.29 & 85.08 & 10.55 \\
F6 & asym-3ms-large & 20 & 3 & 3 & 8.67M & 10.95G & 3.20 & 3.76 & 2.80 & 2.82 & 4.01 & 2.42 & 2.27 & 84.89 & 10.34 \\
\midrule
G1 & asym-3ms-mapping &	20 &3 &3 & 0.625M &	1.54G &	3.10 &	3.74 &	2.70 &	2.70 &	4.01 &	2.28 &	1.91 &	81.85 &	8.19 \\
G2 & asym-6ms-predict &	6 &	6 &	3	& 0.625M &	767.57M &	3.06 &	3.71 &	2.65 & 2.71 &	4.01 &	2.29 &	2.09 &	83.35 &	9.34 \\
\midrule
H1 & FBE-10ms &	- &	- &	10	& 0.656M &	256.1M &	3.15 &	3.68 &	2.72 & 2.80 &	3.93 &	2.36 &	2.05 &	82.98 &	8.91 \\
H2 & FBE-5ms &	- &	- &	5	& 0.644M &	590.6M &	3.11 &	3.65 &	2.68 & 2.75 & 3.93 & 2.32 & 2.00 &	82.90 &	8.60 \\
H3 & FBE-2.5ms &	- &	- &	2.5	& 0.637M &	1.098G &	3.07 &	3.58 &	2.62 & 2.71 & 3.89 &	2.27 & 1.89 & 81.73 & 7.80 \\
\bottomrule
\end{tabular}
}
\end{table*}


\begin{figure}[tb]
    \centering
    \begin{minipage}[b]{0.24\textwidth}
        \centering
        \includegraphics[width=\textwidth,clip,trim=0 5 0 4]{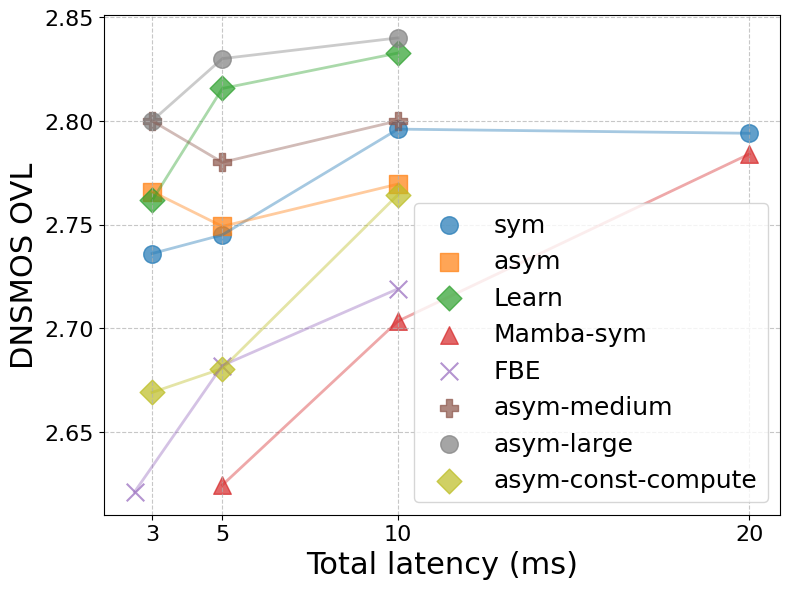}
        \label{fig:image1}
    \end{minipage}
    \hfill
    \begin{minipage}[b]{0.24\textwidth}
        \centering
        \includegraphics[width=\textwidth,clip,trim=0 5 0 4]{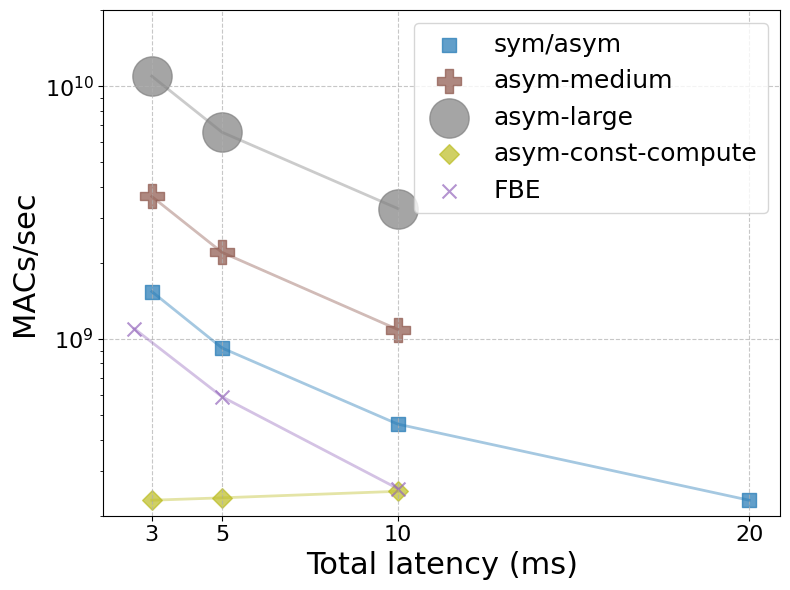}
        \label{fig:image2}
    \end{minipage}
    \caption{Overall results for various low latency methods and model sizes (left), and MACs/sec for models (right).}
    \label{fig:all}
\end{figure}

\section{Experiments}

\subsection{Experimental setup}

We use a large-scale training dataset by mixing speech and noise on-the-fly. 
The speech is 700hour high MOS-rated speech from the LibriVox and AVspeech corpora, while the noise data includes 247-hour recordings from Audioset, Freesound, internal noise sources, and 1 hour of colored stationary noise. 
Except the 65-hour internal noise, all data is available in the 2nd DNS challenge\footnote{https://github.com/microsoft/DNS-Challenge/tree/icassp2021-final}. 
The mixing and reverberation strategies, loss and training scheme are exactly following \cite{braun2021towards}.
As test set, we use a combination of the 4th and 5th DNS Challenge blind test sets\footnote{https://github.com/microsoft/DNS-Challenge}, and an additional simulated test set similarly generated as the training data but from different speech and noise corpora.
To increase the test set difficulty and enhance the significance of the results, we used only lower signal-to-reverberation data portion for DNS test (bottom 50\%, 611 files in total).

As evaluation metrics we use DNSMOS \cite{reddy2021icassp}, a DNN developed to predict mean opinion scores (MOS) for signal (SIG), background (BAK), and overall (OVL) quality. 
Also, we evaluate intrusive objective metrics such as Perceptual Evaluation of Speech Quality (PESQ), Short-Time Objective Intelligibility (STOI), and signal-invariant signal-to-distortion ratio (siSDR) on the validation set.
We also provide the model size (number of parameters), and complexity in terms of multiply-accumulate operations (MACs) measured for processing 1 second of audio data.

\subsection{Experimental results}

\subsubsection{Window type}
From Table~\ref{tab:all} (rows A1-A4, B1-B3, C1-C3, H1-H3) and Figure~\ref{fig:all} (left figure) with confidence intervals (95\%) per model being around 0.017, we can observe the following:
(1). Reducing the latency from 20 to 10 ms does not degrade performance at all for ``sym'', while we see a performance drop for 5~ms and lower.
(2). Surprisingly, the asymmetric window settings perform not significantly different or sometimes mildly worse than symmetric windows.
(3). The learnable transform outperforms non-trainable STFT transforms at higher latency, while the differences seems to become negligible at low latency.
(4). The FBE using overlap-save (H1-H3) performs significantly worse than overlap-add based methods (A2-C3).
The DNSMOS real test results show an inconsistency of ``asym'' gaining performance when going from 5 to 3ms, however the synthetic test set (Table.~\ref{tab:all}) shows an expected constant drop when decreasing latency.

We hypothesize that a powerful model like CRUSE is able to compensate well for smaller analysis windows (symmetric setting), therefore making the asymmetric window technique obsolete.
To investigate this further, we included a weaker model, U-Net, to see if the asymmetric window shines when the model capacity is limited.
Our U-Net is derived by just removing the GRU bottleneck from the CRUSE architecture.
The results for UNet are shown in Table~\ref{tab:unet}.
From Table~\ref{tab:unet} we can observe clear advantages of the asymmetric-window models over the corresponding symmetric-window models, while for the full CRUSE model in Table~\ref{tab:all} the differences are not conclusive or significant. 
This suggests that strong models can compensate for the negative effects of short symmetric windows, while the advantage of asymmetric windows is only significant for weaker models.

\begin{table}[tb]
\centering
\caption{Comparison between asymmetric and symmetric windows. The model architecture is UNet.}
\label{tab:unet}
\adjustbox{width=0.3\textwidth}{
\begin{tabular}{cccccc}
\toprule
\textbf{iWin (ms)} & \textbf{oWin (ms)} & \textbf{SIG} & \textbf{BAK} & \textbf{OVR} \\
\midrule
20 & 20 & 2.80 & 3.74 & 2.47 \\
\midrule
10 & 10 & 2.71 & 3.61 & 2.35 \\
20 & 10 & 2.72 & 3.64 & 2.38 \\
\midrule
5  & 5  & 2.59 & 3.54 & 2.24 \\
20 & 5  & 2.66 & 3.57 & 2.30 \\
\midrule
3  & 3  & 2.49 & 3.53 & 2.16 \\
20 & 3  & 2.62 & 3.52 & 2.25 \\
\bottomrule
\end{tabular}
}
\end{table}

\subsubsection{Model size and computational demand}
When reducing the window size to get lower latency, performance inevitably drops. 
It is however important to note, that reducing the latency by reducing window sizes while keeping the DNN model fixed, increases the compute demand per second of audio data inverse proportionally to the hop-size as shown in Figure~\ref{fig:all} right (blue line), as there is less time to execute the same computation, or the frame rate is higher. 
For reference, if one would keep the compute budget per time constant, the model size or compute demand has to be reduced proportionally to the hop-size. 
We designed several CRUSE models using asymmetric windows for the smaller hop-sizes by changing the number of conv filters to keep the MACs/second roughly constant. 
These models are shown in Figure~\ref{fig:all} as the olive line. 
We can see that the performance drops much more significantly than for the fixed model sizes (orange). 

To investigate how much performance drop caused by reducing the window size can be recovered by increasing model size, we design three model sizes, original, medium, and large, by increasing the number of conv filters.
When enlarging the models (F1-F6 in Table~\ref{tab:all} and Figure~\ref{fig:all}), we observe increasing model sizes can fully compensate for the performance drop of small windows: The large 3 ms latency model performs on par to the original 20 ms latency model.

Lastly, the FBE has lower complexity at same total latency, as its latency only depends on the hop-size, not the (overlapping) window size. 
However, even factoring this advantage in, this model design seems to underperform the other techniques.

\subsubsection{Investigation for Mamba}
The novel Mamba architecture \cite{gu2023mamba} combines a state-space model with a selection mechanism. 
Mamba demonstrated performance on par with or exceeding Transformers across various tasks, including speech enhancement \cite{chao2024investigation} and separation \cite{li2024spmamba}. 
Additionally, Mamba is notable for its computational efficiency.
\textit{However, previous studies \cite{chao2024investigation,li2024spmamba} have not examined Mamba’s performance under low-latency conditions.}
We replace the GRU in CRUSE with Mamba, and the two settings have similar number of model parameters as in Table~\ref{tab:all}.
From Table~\ref{tab:all} (A1, E1-E3), we observe that:
(1). Mamba performs pretty well in standard latency (20 ms).
(2). However, its performance significantly declines when the latency is reduced.


\subsubsection{Investigation for the future-frame prediction technique}
To fully investigate the usefulness of future frame prediction, which is only possible using a signal mapping approach, we compare in Table~\ref{tab:all} the standard filtering based models (A4, B3) with signal mapping based models without prediction (G1) and with prediction (G2) at the same total latency. 
Note that the prediction model (G2) can use a larger window size at the same latency.   
(1). A4 and B3 outperform G1 and G2 across all metrics in both the DNS blind test and simulated test data. The future-frame prediction technique is limited to mapping, which becomes a bottleneck, leading to worse results compared to filtering-based models.
(2). Comparing G2 and G1, G2 performs better on simulated data, consistent with the original paper's findings (which also used simulated data). However, G2 performs worse in the DNS blind test. A possible reason is that the future-frame prediction trick fits the training data well, which is also simulated data, but struggles to generalize on real-world DNS challenge data.

\section{conclusion}
\label{sec:conclusion}
This study addresses the gap in low-latency speech enhancement by evaluating various methods within a unified framework. 
Our findings reveal that asymmetric windows show hardly any benefit over symmetric ones for a strong SE model, but significantly boost the performance of smaller and weaker DNN models. 
The Mamba model excels at standard latency but suffers at lower latencies. 
Learnable windows outperform both asymmetric and symmetric options. The future-frame prediction technique shows no conclusive benefits at same latency and has severe limitations being restricted to lower performing mapping targets. 
Finally, increasing model size enough can fully recover the performance loss associated with reduced window sizes.
We hope the insights and take-aways can help future researchers in developing real-world low-latency SE systems.


\bibliographystyle{IEEEtran}
\bibliography{IEEEabrv,refs}

\end{document}